\documentclass[aps,prl,twocolumn,amsmath,amssymb,superscriptaddress,longbibliography]{revtex4-1}
\usepackage[english]{babel}
\usepackage{xcolor}
\usepackage{amssymb} 
\usepackage{dcolumn}
\usepackage{bm}
\usepackage{graphicx}
\usepackage{amsmath}
\usepackage{braket}
\usepackage[T1]{fontenc}  % ??T1????????????????\k????
\usepackage{CJKutf8}

\usepackage{graphicx}        % standard LaTeX graphics tool

\graphicspath{{pict/}{}}

\usepackage[normalem]{ulem}

\usepackage{dcolumn}%%%%%new
\usepackage{bm}
\usepackage[pdfstartview=FitH, CJKbookmarks=true, bookmarksnumbered=true, bookmarksopen=true, colorlinks=true, pdfborder=001, citecolor=blue, linkcolor=blue, urlcolor=blue, linktocpage=true] {hyperref}

\makeatletter

\begin{document}

\title{Path Complementarity in Hybridized Cavity Polaritons for Directional Quantum Light}

\title{Path Complementarity Enables Directional Quantum Light in Hybridized Cavity Polaritons}

\author{Jing Tang}
%\email{jingtang@gdut.edu.cn}
\affiliation{School of Physics and Optoelectronic Engineering, Guangdong University of Technology, Guangzhou 510006, China}
\affiliation{Guangdong Provincial Key Laboratory of Sensing Physics and System Integration Applications, Guangdong University of Technology, Guangzhou, 510006, China}

\author{Yuangang Deng}
\email{dengyg3@mail.sysu.edu.cn}
\affiliation{Guangdong Provincial Key Laboratory of Quantum Metrology and Sensing $\&$ School of Physics and Astronomy, Sun Yat-Sen University, Zhuhai 519082, China}

\date{\today}
\begin{abstract}
Directional quantum light requires asymmetric photon statistics without sacrificing coherence between propagation paths, a combination that remains difficult to achieve within a single quantum-optical platform. Here we identify a microscopic path-complementarity mechanism in hybridized cavity polaritons that reconciles these two requirements. In a Zeeman-split ring-cavity QED system, cavity backscattering hybridizes the counterpropagating modes, whereas the Zeeman splitting creates spectrally asymmetric excitation channels. Their interplay gives rise, under weak driving, to directional photon blockade coexisting with strong single-photon path entanglement: one input direction exhibits antibunching while the opposite direction is bunched. Remarkably, switching between the two one-photon polariton resonances reverses both the photon-statistical and intensity asymmetries while nearly preserving the path concurrence over a broad parameter regime. Our results establish path complementarity as a general design principle for directional nonclassical light with retained quantum coherence.
\end{abstract}

\date{\today}  

\maketitle 
\section{Introduction}
The ability to control the propagation of quantum light at the single-photon level is a central goal of quantum optics, with direct implications for quantum communication, quantum information processing, and quantum networks~\cite{lodahl2017chiral,sollner2015deterministic,kimble2008quantum,gisin2007quantum,mirhosseini2019cavity}.
Unlike conventional nonreciprocal devices based on macroscopic magnetic biases or engineered dissipative asymmetries~\cite{sounas2017non, metelmann2015nonreciprocal,ruesink2016nonreciprocity}, quantum-optical implementations operate at the level of individual excitations, where propagation direction, photon statistics, and quantum coherence become intrinsically intertwined~\cite{lodahl2017chiral,sollner2015deterministic}. Recent advances in chiral light-matter interactions~\cite{coles2016chirality,hallett2022engineering}, waveguide quantum electrodynamics
~\cite{shen2007strongly,sheremet2023waveguide,roy2017colloquium}, and cavity quantum electrodynamics~\cite{blais2021circuit, haroche2006exploring} have enabled directional photon emission and nonreciprocal quantum transport at single photon level~\cite{sayrin2015nanophotonic,scheucher2016quantum,Shomroni2014all,dong2021all}, through spin-momentum locking~\cite{bliokh2015spin}, synthetic gauge fields~\cite{dalibard2011colloquium}, and time-reversal symmetry breaking~\cite{lu2014topological,ozawa2019topological}.
A fundamental tension nevertheless remains: increasing directional distinguishability generally suppresses coherent superpositions between propagation paths~\cite{englert1996fringe, jakob2007complementarity,PhysRevA.92.012118}. Directionality and path coherence therefore emerge as competing quantum resources.

Cavity quantum electrodynamics provides a natural setting to address this tension because photons and emitters hybridize into polaritonic states whose excitation pathways can interfere coherently~\cite{blais2021circuit,kockum2019ultrastrong,reithmaier2004strong,Liew10}. The resulting nonlinear excitation ladder supports photon blockade, in which a single excitation suppresses subsequent multiphoton absorption through spectral anharmonicity~\cite{imamoglu1997strongly,birnbaum2005,Hennessy07,tang2023strong} or destructive quantum interference~\cite{Liew10,Bamba2011,Tang15, Snijders18, Vaneph18}. More broadly, photon blockade has evolved from a mechanism for generating antibunched photons into a versatile framework for engineering nonclassical light by exploiting additional degrees of freedom, including coupled cavity architectures~\cite{Hartmann06,greentree2006quantum}, multimode interference~\cite{Majumdar2012,Bamba2011}, and hybrid platforms~\cite{Rabl11,wei2026dual}. These developments enable tailored photon statistics and correlations and provide a flexible setting for programmable quantum-light generation~\cite{carusotto2013quantum,volz2012ultrafast,ma2026recent}. In parallel, chiral and nonreciprocal cavity systems have shown that direction-dependent resonance shifts, chiral light-matter couplings, and interference can generate strongly asymmetric photon statistics~\cite{Huang18,li2019nonreciprocal,zheng2025nonreciprocal,graf2022nonreciprocity}. Most existing studies, however, focus on directional transport or photon statistics, whereas the coherence retained between counterpropagating quantum pathways remains comparatively unexplored. This raises a central question: can a single polaritonic structure support strongly directional nonlinear photon statistics while preserving coherent single-photon superpositions between opposite propagation paths?

Here we show that these seemingly competing properties can coexist through path complementarity of hybridized cavity polaritons. We consider a ring-cavity QED system in which backscattering coherently hybridizes two counterpropagating cavity modes coupled to Zeeman-split emitter transitions. These two ingredients play distinct and complementary roles. The Zeeman splitting spectrally differentiates the spin-selective
excitation pathways and generates directional nonlinear response, whereas cavity backscattering coherently hybridizes the
counterpropagating modes and delocalizes the single-photon excitation between them. This interplay is encoded in a microscopic complementarity relation between path predictability and path concurrence, which directly links the photonic composition of each polariton to its directional accessibility.

In the driven-dissipative regime, this microscopic structure yields directional photon blockade coexisting with strong single-photon path entanglement. At a fixed driving frequency, one input direction exhibits pronounced antibunching while the opposite direction is bunched, with finite photon populations in both channels. Resolving the two one-photon polariton resonances, we find that switching between the lower and upper branches reverses both the photon-statistical and intensity asymmetries while nearly preserving the associated path concurrence over a broad range of Zeeman splitting and backscattering strengths. These results establish path complementarity as a design principle for directional nonclassical light: spectral asymmetry selects the propagation direction that accesses the nonlinear polaritonic pathways, whereas coherent mode hybridization preserves path coherence within the single-photon manifold.

\section{Model and hybridized polaritons}

\begin{figure}[ht]
\includegraphics[width=1\columnwidth]{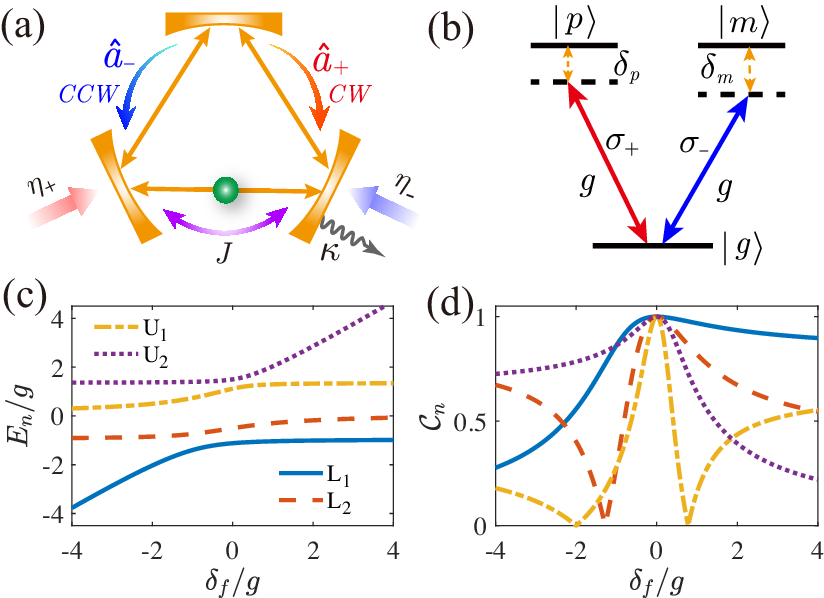}
\caption{(a) Ring-cavity QED system supporting CW and CCW modes $\hat a_+$ and $\hat a_-$, hybridized by backscattering with strength $J$. (b) V-type emitter selectively coupled to the CW and CCW modes through the $\sigma^+$ and $\sigma^-$ transitions, respectively; the excited states are split by $\delta_f$. (c) Single-excitation dressed spectrum versus $\delta_f/g$ for $\Delta_c/g=0$, $\delta_p/g=0.5$, and $J/g=0.5$. (d) Intrinsic photonic path concurrence $\mathcal C_n$ of the four polariton branches. Exchange symmetry gives $\mathcal C_n=1$ at $\delta_f=0$, while finite $\delta_f$ redistributes the CW and CCW weights in a branch-dependent manner.}
\label{model}
\end{figure}

We consider a single three-level emitter coupled to a ring cavity supporting two counterpropagating optical modes, as schematically illustrated in Fig.~\ref{model}(a). The clockwise (CW) and counterclockwise (CCW) cavity modes, denoted by $\hat a_+$ and $\hat a_-$, are coherently hybridized by backscattering with strength $J$~\cite{aoki2006observation,mazzei2007controlled}. The emitter has a V-type level structure with ground state $|g\rangle$ and excited states $|p\rangle$ and $|m\rangle$, selectively coupled to the CW and CCW modes through the $\sigma^+$ and $\sigma^-$ transitions, respectively [Fig.~\ref{model}(b)]. An external magnetic field lifts the degeneracy of the excited states by a Zeeman splitting $\delta_f$, introducing a controllable spectral asymmetry between the two spin-selective light-matter pathways.

In a frame rotating at the drive frequency, the Hamiltonian for input direction $\mu=\pm$ reads
\begin{align}
\hat H^{(\mu)}=&~\Delta_c\left(\hat a_+^\dagger\hat a_++\hat a_-^\dagger\hat a_-\right)
+\delta_p\hat\sigma_{pp} +(\delta_p+\delta_f)\hat\sigma_{mm}  \nonumber\\
&+g\left(\hat a_+^\dagger\hat\sigma_{gp} +\hat a_-^\dagger\hat\sigma_{gm} +{\rm H.c.} \right) \nonumber\\
&+J\left(\hat a_+^\dagger\hat a_- +\hat a_-^\dagger\hat a_+ \right) +i\eta\left( \hat a_\mu^\dagger-\hat a_\mu \right),
\label{eq:H}
\end{align}
where $\Delta_c=\omega_c-\omega_L$ is the cavity detuning, $\delta_p=\omega_p-\omega_L$ is the detuning of the $|g\rangle\leftrightarrow|p\rangle$ transition, $\delta_f=\omega_m-\omega_p$ is the Zeeman splitting between the two excited states, $g$ is the light-matter coupling strength, and $\eta$ is the driving amplitude. We treat $\Delta_c$ as independently tunable and keep $\delta_p$ fixed throughout. The labels $\mu=\pm$ denote excitation from opposite propagation directions. Importantly, the input direction does not create the asymmetry; it probes an intrinsic polariton structure set by two distinct mechanisms. Backscattering $J$ hybridizes the counterpropagating modes, whereas the Zeeman splitting $\delta_f$ lifts the spectral equivalence of the two spin-selective excitation pathways.

The dissipative dynamics is described by the master equation
\begin{align}
\frac{d\hat\rho}{dt}=&-i[\hat H^{(\mu)},\hat\rho]+\kappa\mathcal D[\hat a_+]\hat\rho+\kappa\mathcal D[\hat a_-]\hat\rho   \nonumber\\
&+\gamma\mathcal D[\hat\sigma_{gp}]\hat\rho +\gamma\mathcal D[\hat\sigma_{gm}]\hat\rho,
\label{eq:master}
\end{align}
where $\mathcal D[\hat o]\hat\rho=\hat o\hat\rho\hat o^\dagger -\left(\hat o^\dagger\hat o\hat\rho +\hat\rho\hat o^\dagger\hat o \right)/2$ is the standard Lindblad dissipator.  We focus on the weak-drive regime, where the microscopic origin of the directional response is already encoded in the single-excitation polaritons of the undriven system.

The ingredients of Eq.~(\ref{eq:H}) are available in current microresonator and nanophotonic QED platforms. Strong coupling between single atoms and whispering-gallery modes has been demonstrated experimentally~\cite{aoki2006observation,junge2013strong}. The polarization structure of tightly confined optical fields enables spin-selective coupling of circular atomic transitions to opposite propagation directions~\cite{junge2013strong}. The Zeeman splitting between the two circular transitions can be continuously tuned by an external magnetic field. Controlled scattering provides coherent coupling between counterpropagating cavity resonances~\cite{mazzei2007controlled}. Single-atom microresonators have also demonstrated photon-number-dependent transport, single-photon routing, and quantum-optical circulation~\cite{dayan2008photon,Shomroni2014all,scheucher2016quantum}.%%%%%%%%%%%%%%%%%%%%%%%%%

{\em Hybridized dressed polaritons}.---To identify the microscopic structure underlying the directional response, we first analyze the intrinsic polariton structure independently of the input direction. In the weak-drive regime, the relevant structure is already captured by the single-excitation manifold of the undriven Hamiltonian in Eq.~(\ref{eq:H}). In the single-excitation manifold, we use the ordered basis ${|g,1_+,0_-\rangle, |g,0_+,1_-\rangle ,|p,0_+,0_-\rangle, |m,0_+,0_-\rangle}$, where the last two entries denote CW and CCW photon numbers. The projected Hamiltonian is
\begin{equation}
\mathcal H^{(1)}=\begin{pmatrix}
\Delta_c & J & g & 0\\
J & \Delta_c & 0 & g\\
g & 0 & \delta_p & 0\\
0 & g & 0 & \delta_p+\delta_f
\end{pmatrix},
\label{eq:H1}
\end{equation}
Its eigenstates are the single-excitation polaritons
\begin{align}
|\Psi_n\rangle=&~c_{n,+}|g,1_+,0_-\rangle +c_{n,-}|g,0_+,1_-\rangle       \nonumber\\
&+ c_{n,p}|p,0_+,0_-\rangle +c_{n,m}|m,0_+,0_-\rangle ,
\label{eq:dressed}
\end{align}
with $\mathcal H^{(1)}|\Psi_n\rangle=E_n|\Psi_n\rangle$.  Here $n=1,\ldots,4$ labels the four single-excitation polariton branches.

The effect of backscattering is most transparent in the symmetric limit $\delta_f=0$. Introducing the symmetric and antisymmetric photonic states
\begin{align}
|C_{s/a}\rangle
&=
\frac{|g,1_+,0_-\rangle\pm|g,0_+,1_-\rangle}{\sqrt2},
\nonumber\\
|P_{s/a}\rangle
&=
\frac{|p,0_+,0_-\rangle\pm|m,0_+,0_-\rangle}{\sqrt2}.
\end{align}
The Hamiltonian decomposes into independent symmetric and antisymmetric
polaritonic sectors,
\[
\mathcal H_{\xi}
=
\begin{pmatrix}
\Delta_c+s_{\xi}J & g\\
g & \delta_p
\end{pmatrix},
\qquad
s_s=+1,\quad s_a=-1.
\]
The corresponding eigenenergies are
\begin{equation}
E_{\xi}^{\lambda}
=
\frac{\Delta_c+s_{\xi}J+\delta_p}{2}
+
\lambda
\sqrt{
g^2+
\left(
\frac{\Delta_c+s_{\xi}J-\delta_p}{2}
\right)^2
},
\label{eq:Esa}
\end{equation}
where $\xi=s,a$ and $\lambda=\pm$ labels the upper and lower polariton
branches.

Equation~(\ref{eq:Esa}) shows that backscattering splits the symmetric and antisymmetric polariton sectors. At $\delta_f=0$, the Hamiltonian is invariant under the simultaneous exchange $\hat a_+\leftrightarrow\hat a_-$ and $|p\rangle\leftrightarrow|m\rangle$. Consequently, each polariton has equal CW and CCW photonic weights, $|c_{n,+}|=|c_{n,-}|$. The two opposite input configurations therefore access symmetry-related polaritonic excitation pathways.

A finite Zeeman splitting breaks this exchange symmetry by making the two optical transitions spectrally inequivalent. In ${|C_s\rangle,|C_a\rangle,|P_s\rangle,|P_a\rangle}$ basis, Eq.~(\ref{eq:H1}) takes the form
\begin{equation}
\mathcal H^{(1)}=\begin{pmatrix}
\Delta_c+J &0&g&0\\
0&\Delta_c-J&0&g\\
g&0&\bar\delta&-\delta_f/2\\
0&g&-\delta_f/2&\bar\delta
\end{pmatrix}, \nonumber
\end{equation}
with $\bar\delta=\delta_p+{\delta_f}/{2}.$ This representation makes the symmetry-breaking mechanism explicit. Backscattering $J$ splits the symmetric and antisymmetric photonic sectors, whereas the Zeeman term $-\delta_f/2$ mixes their atomic components. Finite $\delta_f$ therefore mixes polaritonic sectors that are decoupled at $\delta_f=0$, reconstructing both their eigenenergies and CW-CCW composition. The resulting branch-dependent reconstruction is evident in the dressed spectrum of Fig.~\ref{model}(c).

For a given polariton, Eq.~(\ref{eq:H1}) gives $c_{n,p}={g}c_{n,+}/({E_n-\delta_p})$ and $c_{n,m}={g}c_{n,-}/({E_n-\delta_p-\delta_f})$. Eliminating the atomic amplitudes yields the ratio of the two photonic components
\begin{equation}
\frac{c_{n,-}}{c_{n,+}}= \frac{(E_n-\Delta_c) (E_n-\delta_p)  -{g^2}} {J({E_n-\delta_p})},
\label{eq:ratio}
\end{equation}
Equation~(\ref{eq:ratio}) directly links the dressed spectrum to the CW-CCW composition of each polariton. This connection acquires a direct operational meaning in the two-sided excitation protocol. For the vacuum state $|G\rangle=|g,0_+,0_-\rangle$, the excitation amplitude of the $n$th dressed polariton from input $\mu=\pm$ is
\begin{equation}
\Omega_n^{(\mu)}=\langle\Psi_n|\hat H_d^{(\mu)}|G\rangle= i\eta c^*_{n,\mu},
\label{eq:Omega}
\end{equation}
where $\hat H_d^{(\mu)}=i\eta(\hat a_\mu^\dagger-\hat a_\mu)$ is the drive term in Eq.~(\ref{eq:H}). Thus, the intrinsic polaritons are input independent, whereas their directional accessibility is set directly by their CW and CCW photonic weights. For $\delta_f=0$, the eigenvalue condition $E_n-\Delta_c-g^2/(E_n-\delta_p)=\pm J$ gives
$c_{n,-}/c_{n,+}=\pm1$ through Eq.~(\ref{eq:ratio}), and hence
$|\Omega_n^{(+)}|=|\Omega_n^{(-)}|$; finite $\delta_f$ lifts this
constraint and produces branch-dependent directional accessibility.

{\em Microscopic path complementarity}.---The photonic amplitudes $c_{n,\pm}$ simultaneously determine the directional accessibility of a polariton and the distribution of its single-photon weight between the two propagation paths. To expose this connection, we introduce the amplitude ratio $R_n \equiv | {c_{n,-}}/{c_{n,+}}|$ which follows directly from Eq.~(\ref{eq:ratio}). The normalized single-photon component of the $n$th polariton can then be written as
\begin{equation}
|\psi_n^{\rm ph}\rangle=\frac{|1_+,0_-\rangle+R_ne^{i\phi_n}|0_+,1_-\rangle} {\sqrt{1+R_n^2}},
\label{eq:phstate}
\end{equation}
where $\phi_n$ is the relative phase between the two propagation amplitudes.

Following the standard two-path formulation~\cite{jaeger1995two,englert1996fringe}, we define the path predictability $\mathcal P_n={|1-R_n^2|}/({1+R_n^2})$, which measures the imbalance between the CW and CCW photonic weights. Thus, $\mathcal P_n=0$ corresponds to equal path weights, whereas $\mathcal P_n=1$ denotes complete localization in one propagation path.

Using $|\Omega_n^{(\pm)}|=\eta|c_{n,\pm}|$, the predictability can equivalently be written as
\begin{equation}
\mathcal P_n=\frac{\left| |\Omega_n^{(+)}|^2- |\Omega_n^{(-)}|^2 \right|}{|\Omega_n^{(+)}|^2+|\Omega_n^{(-)}|^2},
\label{eq:Pomega}
\end{equation}
Thus, $\mathcal P_n$ has a direct spectroscopic meaning: it measures the imbalance with which a given polariton is accessed from opposite input directions. The complementary quantity is the intrinsic photonic path concurrence: $\mathcal C_n={2R_n}/({1+R_n^2})$, which equals unity for an equal coherent path superposition and vanishes for complete localization. Equivalently, $\mathcal C_n={2|c_{n,+}c_{n,-}|}/({|c_{n,+}|^2+|c_{n,-}|^2})$.

Because Eq.~(\ref{eq:phstate}) is a pure two-path state, the predictability and concurrence obey the exact relation
\begin{equation}
\mathcal P_n^2+\mathcal C_n^2=1.
\label{eq:Complementarity}
\end{equation}
The significance of Eq.~(\ref{eq:Complementarity}) lies in its microscopic embedding in the hybridized polariton structure, rather than in the two-path identity itself. Because the same amplitudes $c_{n,\pm}$  determine both the excitation strengths $|\Omega_n^{(\pm)}|$ and the path concurrence, the polariton wave function simultaneously fixes its directional accessibility and its internal path coherence. At $\delta_f=0$, exchange symmetry enforces $R_n=1$, yielding $\mathcal P_n=0$ and $\mathcal C_n=1$ for all four polariton branches. Finite Zeeman splitting reconstructs the polariton wave functions and redistributes the CW and CCW weights in a branch-dependent manner.

This redistribution is evident in Fig.~\ref{model}(d). Some polariton branches become strongly path polarized, with $\mathcal C_n$ approaching zero, whereas others retain substantial intrinsic path concurrence at finite $\delta_f$. The Zeeman splitting therefore does not simply suppress path coherence; instead, it redistributes coherence among the hybridized polariton branches through their reconstructed CW-CCW composition.

The relation in Eq.~(\ref{eq:Complementarity}) is intrinsic to the normalized single-photon component of a pure polariton eigenstate. Directional photon blockade, by contrast, is a nonlinear driven-dissipative effect and need not obey this pure-state identity. We now examine how this intrinsic polariton structure is reflected in the driven steady state.
%%%%%%%%%%%%%%%%%%%%%%

\section{Results}
{\em Directional photon blockade and path entanglement}.---We now turn from the intrinsic complementarity of individual polaritons to the driven-dissipative steady state. For each input direction $\mu=\pm$, we obtain the steady state $\rho_s^{(\mu)}$ from Eq.~(\ref{eq:master}) and characterize the driven mode by its photon population $n_\mu={\rm Tr}\!\left(\hat a_\mu^\dagger\hat a_\mu\rho_s^{(\mu)}\right)$
and equal-time second-order correlation $g_\mu^{(2)}(0)={{\rm Tr}\!\left(\hat a_\mu^{\dagger 2}\hat a_\mu^2\rho_s^{(\mu)}\right)}/{n_\mu^2},$ which quantify the intensity and photon statistics, respectively~\cite{carmichael2013statistical}. For a monitored output port with vacuum input, the normally ordered normalized correlations are directly inherited by the output field through the standard input-output relation~\cite{flayac2013input}.

To quantify the path entanglement carried by the single-photon component, we trace out the emitter and condition the photonic steady state on the single-photon manifold,
$\rho_1^{(\mu)}=\frac{\Pi_1\,{\rm Tr}_{\rm at}\!\left[\rho_s^{(\mu)}\right]\Pi_1}{{\rm Tr}\!\left\{\Pi_1\,{\rm Tr}_{\rm at}\!\left[\rho_s^{(\mu)}\right]\right\}}$, where $\Pi_1=|1_+,0_-\rangle\langle1_+,0_-|+|0_+,1_-\rangle\langle0_+,1_-|.$ The corresponding conditional single-photon path concurrence is 
$C_\mu=2\left| \langle1_+,0_-| \rho_1^{(\mu)} |0_+,1_-\rangle \right|$, which measures the CW-CCW mode entanglement conditioned on the presence of a single intracavity photon~\cite{wootters1998entanglement,hessmo2004experimental}.

The intrinsic concurrence $\mathcal C_n$ and the conditional concurrence $C_\mu$ represent the same path-concurrence measure evaluated for different states. The former refers to the normalized single-photon component $|\psi_n^{\rm ph}\rangle$ of an isolated polariton eigenstate, whereas the latter is evaluated for the conditional state $\rho_1^{(\mu)}$ of the driven-dissipative system. When a single well-resolved polariton $n$ dominates the response to input $\mu$, $\rho_1^{(\mu)}\simeq|\psi_n^{\rm ph}\rangle\langle\psi_n^{\rm ph}|$ and $C_\mu\simeq\mathcal C_n$.

\begin{figure}[ht]
\includegraphics[width=0.9\columnwidth]{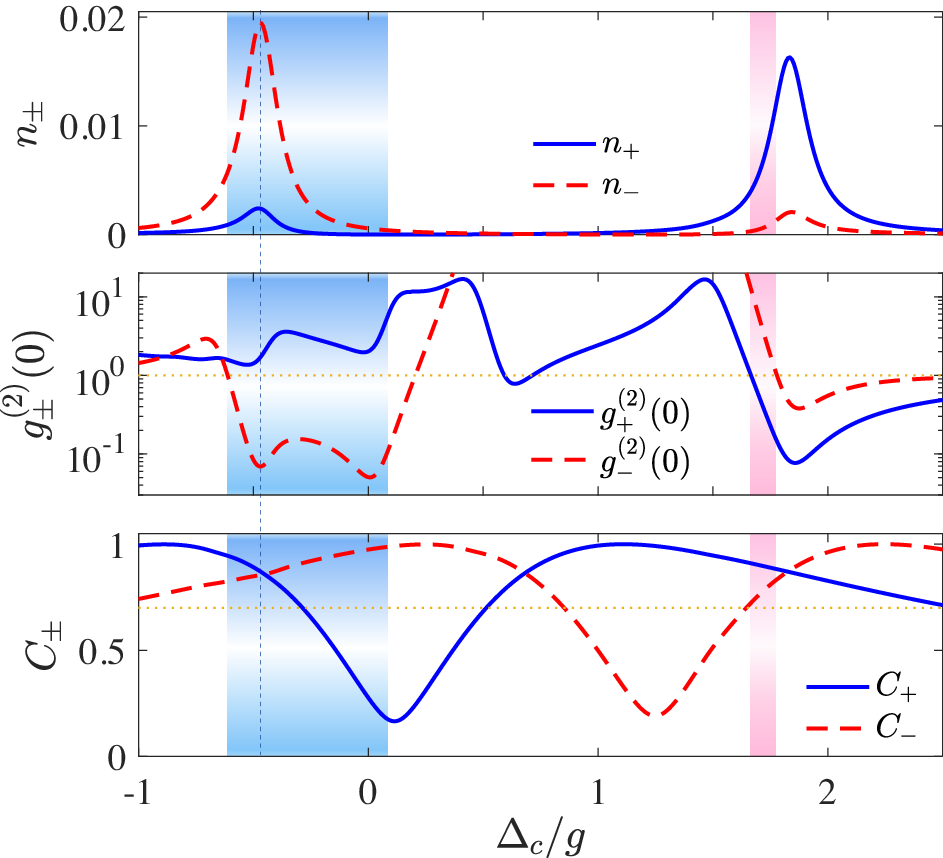}
\caption{Directional photon statistics and single-photon path entanglement. (a) Steady-state photon populations $n_+$ and $n_-$, (b) equal-time second-order correlations $g_+^{(2)}(0)$ and $g_-^{(2)}(0)$, and (c) conditional single-photon path concurrences $C_+$ and $C_-$ as functions of the cavity detuning $\Delta_c$. The horizontal dotted lines in (b) and (c) indicate $g^{(2)}(0)=1$ and $C_\pm=0.7$, respectively. The vertical dotted line marks a representative operating point exhibiting directional antibunching together with strong path entanglement. The other parameters are $g/\kappa=6$, $J/g=1$, $\delta_f/g=8$, $\delta_p/g=0.8$, $\gamma/\kappa=0.1$, and $\eta/\kappa=0.1$.}
\label{fig2}
\end{figure}

Figure~\ref{fig2} shows the photon populations, second-order correlations, and conditional path concurrences versus $\Delta_c$ for $\delta_f/g=8$ and $J/g=1$. Although opposite inputs probe the same intrinsic polariton spectrum, their driven responses become strongly asymmetric at finite Zeeman splitting. The origin of this asymmetry is already visible in the single-excitation manifold: Eq.~(\ref{eq:Omega}) gives $|\Omega_n^{(\pm)}|=\eta|c_{n,\pm}|$. The Zeeman splitting redistributes the CW and CCW weights among the polariton branches, making a given branch unequally accessible from opposite inputs. When propagated into the nonlinear excitation manifold, this directional accessibility produces pronounced asymmetries in both the photon populations [Fig.~\ref{fig2}(a)] and photon statistics [Fig.~\ref{fig2}(b)].

Two detuning windows are particularly notable: the opposite inputs exhibit markedly different photon statistics, with their directional roles interchanged between the two resonances.  At $\Delta_c/g\simeq-0.48$, for example, the $-$ input exhibits strong antibunching, $g_-^{(2)}(0)\simeq6.2\times10^{-2}$, whereas the $+$ input is bunched, $g_+^{(2)}(0)\simeq1.62$. The corresponding photon populations remain finite, with $n_-\simeq2.1\times10^{-2}$ and $n_+\simeq2.6\times10^{-3}$. At $\Delta_c/g\simeq1.75$, the directional asymmetry reverses: $g_+^{(2)}(0)\simeq0.20$ while $g_-^{(2)}(0)\simeq1.76$, with $n_+\simeq1.0\times10^{-2}$ and $n_-\simeq1.0\times10^{-3}$. Thus, the input direction controls not only the mean intensity but also which excitation pathway supports antibunched emission.

Remarkably, the directional reversal of the photon statistics does not require localization of the conditional single-photon state. At $\Delta_c/g\simeq-0.48$, both conditional concurrences remain large, $C_-\simeq0.85$ and $C_+\simeq0.88$. At the reversed operating point $\Delta_c/g\simeq1.75$, they remain $C_+\simeq0.89$ and $C_-\simeq0.81$. Thus, strong directional photon statistics coexist with substantial CW-CCW path entanglement in the conditional single-photon state.

This coexistence follows from the distinct roles of $\delta_f$ and $J$. The Zeeman splitting $\delta_f$ breaks the spectral equivalence of the spin-selective light-matter pathways and generates branch-dependent directional accessibility, whereas backscattering $J$ hybridizes the counterpropagating modes and maintains coherent path mixing. Their interplay allows strong directional photon statistics to coexist with substantial conditional path concurrence.

Figure~\ref{fig2} thus provides the driven-dissipative manifestation of the single-polariton complementarity: directional photon statistics
and path coherence are rooted in the same hybridized CW-CCW polariton structure, yet remain distinct properties of the quantum response. We next resolve the response along the two one-photon polariton branches and show that switching between them reverses the directional photon
statistics while leaving the associated path concurrence nearly unchanged.
%%%%%%%%%%%%%%%%%%%%%%%%%%%%

{\em Dressed-resonance reversal of directional response}.---We next trace the two directional-response windows in Fig.~\ref{fig2} to the underlying one-photon polariton resonances. At resonance, eliminating the atomic amplitudes from the zero-quasienergy single-excitation problem gives the effective photonic Hamiltonian
\begin{equation}
{\cal H}_{\rm ph}^{\rm eff}=\begin{pmatrix}
\Delta_c-\chi_p & J\\
J & \Delta_c-\chi_m
\end{pmatrix},
\label{Heff_ph}
\end{equation}
where $\chi_p={g^2}/{\delta_p}$ and $\chi_m={g^2}/({\delta_p+\delta_f})$ denote the emitter-induced self-energy shifts of the two spin-selective photonic channels at zero quasienergy. The lower ($\mathrm L$) and upper ($\mathrm U$) one-photon resonances then follow from $\det{\cal H}_{\rm ph}^{\rm eff}=0$:
\begin{align}
\Delta_{c,\mathrm L}^{\rm res}&=\frac{\chi_p+\chi_m}{2}-\sqrt{J^2+\left(\frac{\chi_p-\chi_m}{2} \right)^2},    \nonumber\\
\Delta_{c,\mathrm U}^{\rm res}&=\frac{\chi_p+\chi_m}{2}+\sqrt{J^2+\left(\frac{\chi_p-\chi_m}{2} \right)^2},
\label{eq:resonance}
\end{align}
These lossless resonance branches provide the natural spectral reference for the full driven-dissipative response. For $J=0$ and $\chi_m<\chi_p$, Eq.~(\ref{eq:resonance}) reduces to $\Delta_{c,\mathrm L}^{\rm res}=g^2/(\delta_p+\delta_f)$ and $\Delta_{c,\mathrm U}^{\rm res}=g^2/\delta_p$. Finite $J$ coherently hybridizes the two propagation channels and separates the resulting dressed resonances.

\begin{figure}[ht]
\includegraphics[width=1\columnwidth]{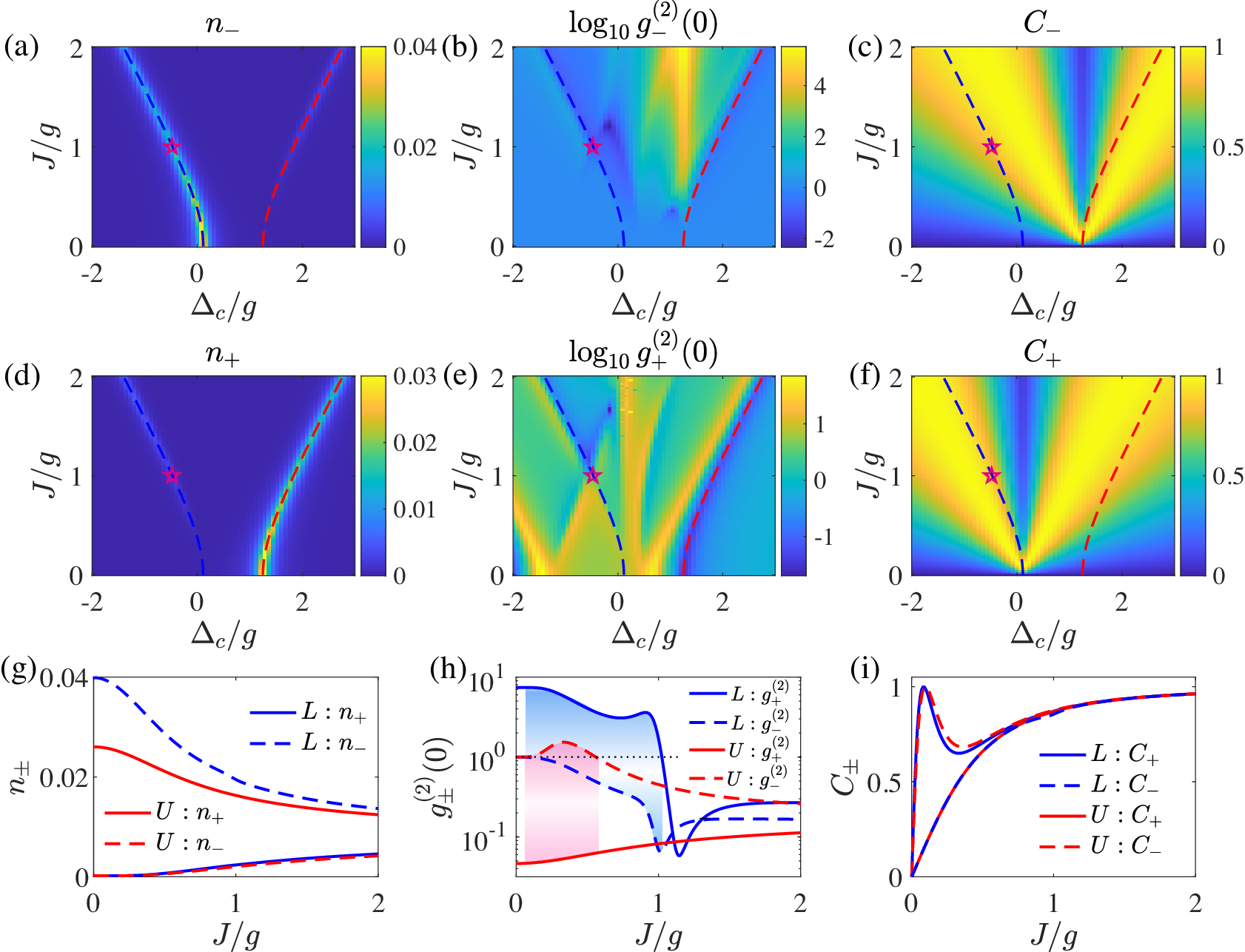}
\caption{Branch-resolved directional photon statistics and path entanglement. (a)-(c) Photon population $n_-$, equal-time correlation $\log_{10}g_-^{(2)}(0)$, and conditional single-photon path concurrence $C_-$ for the $-$ input in $(\Delta_c,J)$ plane. (d)-(f) Corresponding quantities for the $+$ input. Blue and red dashed curves denote the lower and upper one-photon resonances, respectively. The stars mark the lower-branch operating
point at $J/g=1$ and $\Delta_c/g\simeq-0.48$, corresponding to the vertical dotted line in Fig.~\ref{fig2}. (g)-(i) $n_\pm$, $g_\pm^{(2)}(0)$, and $C_\pm$, evaluated along the lower (blue) and upper (red) resonance branches. Solid and dashed curves denote the $+$ and $-$ inputs, respectively. The horizontal dotted line in (h) marks $g^{(2)}(0)=1$, and the shaded regions indicate the intervals of opposite photon statistics, $[g_+^{(2)}(0)-1][g_-^{(2)}(0)-1]<0$. Here $\delta_f/g=8$ and $\delta_p/g=0.8$; the other parameters are the same as in Fig.~\ref{fig2}.}
\label{fig3}
\end{figure}

The connection between the dressed resonances and the full quantum response is made explicit in Figs.~\ref{fig3}(a)--\ref{fig3}(f), which show the direction-resolved photon population, equal-time correlation, and conditional path concurrence in the $(\Delta_c,J)$ plane, with $\Delta_{c,\mathrm L}^{\rm res}$ and $\Delta_{c,\mathrm U}^{\rm res}$ superimposed as the blue and red dashed curves, respectively. The dominant response features closely track these branches, showing that the one-photon polariton spectrum organizes the driven-dissipative response. The stars mark the lower-branch operating point at $J/g=1$ and $\Delta_c/g\simeq-0.48$, corresponding to the strongly directional response highlighted by the vertical dotted line in Fig.~\ref{fig2}.

To reveal the branch-dependent directional response more directly, Figs.~\ref{fig3}(g)-\ref{fig3}(i) show the corresponding observables evaluated along the two dressed resonances. The populations in Fig.~\ref{fig3}(g) already exhibit a pronounced branch-dependent directional imbalance. The population imbalance mainly reflects the direction-dependent coupling to the resonant polariton, whereas $g^{(2)}(0)$ further probes the nonlinear response beyond the single-excitation manifold.

The resulting photon statistics are shown in Fig.~\ref{fig3}(h). Along the lower resonance $\Delta_{c,\mathrm L}^{\rm res}$, the two input directions develop strongly different correlations, with one direction becoming antibunched while the opposite direction remains on the bunched side of the Poissonian boundary over an extended interval of $J$. Along the upper resonance $\Delta_{c,\mathrm U}^{\rm res}$, this directional character is reversed. The shaded regions highlight the intervals satisfying $[g_+^{(2)}(0)-1][g_-^{(2)}(0)-1]<0$, demonstrating that switching between the two dressed resonances reverses which input direction supports antibunched emission.

This branch reversal identifies the two windows in Fig.~\ref{fig2} as the lower and upper members of the same hybridized polariton structure. The Zeeman splitting makes the two spin-selective channels spectrally inequivalent, whereas $J$ coherently mixes their photonic components. Switching between the two resonance branches therefore interchanges their directional accessibility and reverses the observed photon-statistical asymmetry. The resonance branch thus provides a spectral control knob for reversing the direction of nonclassical photon emission.

In contrast to the photon-statistical reversal, the conditional path concurrence displays a nearly branch-exchange symmetry.  As shown in Fig.~\ref{fig3}(i), $C_-$ on the lower branch nearly coincides with $C_+$ on the upper branch, while $C_+$ on the lower branch similarly follows $C_-$ on the upper branch. To expose the microscopic origin of this correspondence, we define the effective photonic detunings at the two resonances as $D_{p,\nu}=\Delta_{c,\nu}^{\rm res}-\chi_p$ and $D_{m,\nu}=\Delta_{c,\nu}^{\rm res}-\chi_m$ with $\nu=\mathrm L,\mathrm U.$
Eq.~(\ref{eq:resonance}) gives the branch-exchange relations $D_{p,\mathrm U}=-D_{m,\mathrm L}$ and $D_{m,\mathrm U}=-D_{p,\mathrm L}$. Thus, switching the resonance branch exchanges the two effective channel detunings up to a sign.

The same exchange structure appears directly in the intrinsic polariton wave functions. At either resonance, Eq.~(\ref{eq:ratio}) gives
\begin{equation}
R_\nu^{\rm res}=\left| \frac{c_{\nu,-}}{c_{\nu,+}} \right| =\left| \frac{D_{p,\nu}}{J} \right|.
\label{eq:Rres}
\end{equation}
Meanwhile, the resonance condition $\det{\cal H}_{\rm ph}^{\rm eff}=0$  implies $D_{p,\nu}D_{m,\nu}=J^2$. This further yields the exact relation
$R_{\mathrm U}^{\rm res}={1}/{R_{\mathrm L}^{\rm res}}$.

Since $\mathcal C_n=2R_n/(1+R_n^2)$ is invariant under
$R_n\rightarrow1/R_n$, the reciprocal relation
$R_{\mathrm U}^{\rm res}=1/R_{\mathrm L}^{\rm res}$ immediately yields
\begin{equation}
\mathcal C_{\mathrm L}^{\rm res}
=
\mathcal C_{\mathrm U}^{\rm res}
=
\frac{J}
{\sqrt{J^2+(\chi_p-\chi_m)^2/4}},
\label{eq:Cintrinsic_exchange}
\end{equation}
At the same time, their normalized CW and CCW weights are interchanged. Branch switching therefore reverses the directional accessibility while exactly preserving the intrinsic path concurrence.

This exact single-polariton relation explains the approximate branch-direction correspondence of the driven steady state,
\begin{equation}
\begin{aligned}
C_-\!\left(\Delta_{c,\mathrm L}^{\rm res}\right) &\simeq C_+\!\left(\Delta_{c,\mathrm U}^{\rm res}\right), \\
C_+\!\left(\Delta_{c,\mathrm L}^{\rm res}\right) &\simeq C_-\!\left(\Delta_{c,\mathrm U}^{\rm res}\right).
\end{aligned}
\label{C_branch_direction}
\end{equation}
Equation~(\ref{C_branch_direction}) is directly supported by the full driven-dissipative calculations in Fig.~\ref{fig3}(i), where the concurrence curves exhibit a clear pairwise correspondence under simultaneous exchange of resonance branch and input direction. The correspondence is not exact because $C_\mu$ is evaluated for the conditional driven steady state and therefore includes dissipation, finite-drive mixing, and contributions beyond an isolated polariton. Increasing $J$ strengthens CW-CCW hybridization and generally enhances the conditional path concurrence. Crucially, this enhanced path entanglement does not eliminate the directional photon-statistical contrast in Fig.~\ref{fig3}(h). An extended regime therefore exists in which strongly asymmetric photon statistics coexist with substantial single-photon path entanglement.

The dressed-resonance exchange thus reverses the directional photon statistics while approximately preserving the corresponding conditional path concurrence upon exchange of the input direction. The exact intrinsic relation in Eq.~(\ref{eq:Cintrinsic_exchange}) and its driven-state counterpart in Eq.~(\ref{C_branch_direction}) connect the microscopic polariton complementarity directly to the observable branch-resolved response. We finally examine the robustness of this coexistence over the broader $(J,\delta_f)$ parameter space.
%%%%%%%%%%%%%%%%%%%%%%%%%%%%%%

\begin{figure}[ht]
\includegraphics[width=1\columnwidth]{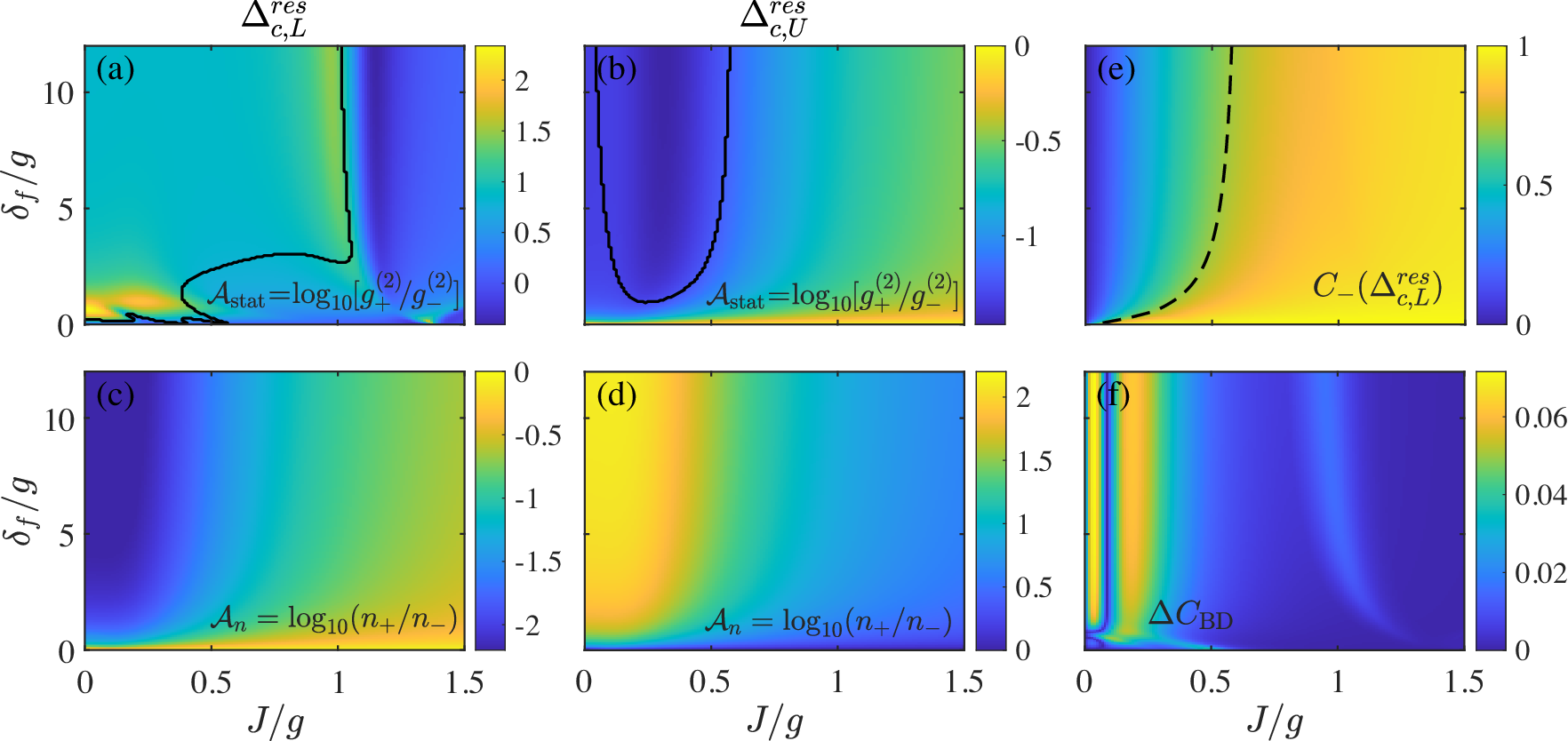}
\caption{Branch-resolved directional quantum response in the $(J,\delta_f)$ parameter plane. The first and second columns correspond to the lower ($\mathrm L$) and upper ($\mathrm U$) one-photon resonance branches $\Delta_{c,\mathrm L}^{\rm res}$ and $\Delta_{c,\mathrm U}^{\rm res}$, respectively. (a),(b) Signed photon-statistical asymmetry $\mathcal A_{\rm stat}=\log_{10}[g_+^{(2)}(0)/g_-^{(2)}(0)]$. Black solid contours delimit the regions of opposite photon statistics, $[g_+^{(2)}(0)-1][g_-^{(2)}(0)-1]<0$. (c),(d) Signed intensity asymmetry $\mathcal A_n=\log_{10}(n_+/n_-)$. (e) Conditional single-photon path concurrence $C_-(\Delta_{c,\mathrm L}^{\rm res})$; the black dashed contour marks $C_-=0.7$. (f) Worst-case branch-direction concurrence mismatch $\Delta C_{\rm BD}$. Branch switching reverses both photon-statistical and intensity asymmetry while approximately preserving the branch-direction concurrence correspondence. Here $\delta_p/g=0.8$; other parameters are as in Fig.~\ref{fig2}.}
\label{fig4}
\end{figure}

%%%%%%%%%%%%%
{\em Robust branch-direction correspondence}.---We finally examine the robustness of the branch-dependent directional response over the $(J,\delta_f)$ plane, evaluating all observables at the lower and upper one-photon resonances $\Delta_{c,\mathrm L}^{\rm res}$ and $\Delta_{c,\mathrm U}^{\rm res}$. To resolve the direction of the photon-statistical asymmetry, we introduce the signed asymmetry $\mathcal A_{\rm stat}=\log_{10} \frac{g_+^{(2)}(0)}{g_-^{(2)}(0)}.$ Its sign identifies the direction with the smaller second-order correlation: $\mathcal A_{\rm stat}>0$ corresponds to $g_-^{(2)}(0)<g_+^{(2)}(0)$, whereas $\mathcal A_{\rm stat}<0$ denotes the opposite ordering. Figures~\ref{fig4}(a) and~\ref{fig4}(b) show a pronounced sign reversal of $\mathcal A_{\rm stat}$ between the two resonance branches. To distinguish opposite photon statistics from a simple difference in correlation strength, the black contours delimit the regions: $[g_+^{(2)}(0)-1] [g_-^{(2)}(0)-1]<0,$ where one input is antibunched and the other bunched. The extended enclosed regions show that the branch-induced reversal persists over broad ranges of both $\delta_f$ and $J$, rather than being confined to isolated operating points.

A similar reversal appears in the intensity response. Figures~\ref{fig4}(c) and~\ref{fig4}(d) show the signed intensity asymmetry $\mathcal A_n=\log_{10}(n_+/n_-),$ whose sign also reverses between the lower and upper branches over most of the parameter plane. The two asymmetries probe distinct properties: $\mathcal A_{\rm stat}$ measures the directional contrast in photon statistics, whereas $\mathcal A_n$ measures the intensity imbalance. Their simultaneous sign reversal shows that branch switching reverses both the statistical and intensity asymmetries.

Crucially, strong directional photon-statistical asymmetry does not require localization of the conditional single-photon state. Figure~\ref{fig4}(e) shows $C_-(\Delta_{c,\mathrm L}^{\rm res})$, with the dashed contour marking $C_-=0.7$. Comparison with Fig.~\ref{fig4}(a) reveals a substantial overlap between the opposite-statistics region and the region of large path concurrence, particularly for intermediate backscattering $J/g\simeq0.55$-$1$. This overlap defines an extended coexistence regime of strong directional photon-statistical contrast and substantial single-photon path entanglement. In this regime, $J$ is large enough to maintain coherent CW-CCW hybridization while preserving the branch-dependent directional accessibility.

We next quantify how accurately the branch-direction concurrence correspondence survives across the parameter plane. Motivated by the branch-direction correspondence in Eq.~(\ref{C_branch_direction}), we define the worst-case branch-direction mismatch $\Delta C_{\rm BD}
=\max\left\{ \left| C_-\!\left(\Delta_{c,\mathrm L}^{\rm res}\right) - C_+\!\left(\Delta_{c,\mathrm U}^{\rm res}\right) \right|, \left| C_+\!\left(\Delta_{c,\mathrm L}^{\rm res}\right) - C_-\!\left(\Delta_{c,\mathrm U}^{\rm res}\right) \right| \right\}.$ As shown in Fig.~\ref{fig4}(f), $\Delta C_{\rm BD}$ remains small over most of the parameter plane, with appreciable deviations confined to limited regions. Thus, the branch-direction correspondence remains accurate over broad ranges of $J$ and $\delta_f$, including regions with strong directional photon-statistical contrast. The correspondence found in Fig.~\ref{fig3}(i) therefore extends well beyond the representative parameter cut.

Taken together, Fig.~\ref{fig4} establishes a broad parameter regime in which directional photon statistics and single-photon path entanglement coexist. The resonance branch selects the favored input direction, whereas backscattering maintains coherent mixing of the CW and CCW components. Their interplay allows opposite photon statistics to coexist with substantial path entanglement, while branch switching reverses the directional response with only a small change in the corresponding concurrence. Together with the intrinsic relation $\mathcal P_n^2+\mathcal C_n^2=1$ and the branch-exchange relation established above, these results identify path complementarity as the organizing principle linking hybridized cavity polaritons to directional nonclassical light.
%%%%%%%%%%%%%%%%%%%%%%%%%%%%%%%%%%

\section{CONCLUSION}
In conclusion, we have identified path complementarity as a microscopic organizing principle for directional quantum light in hybridized cavity polaritons. The same CW-CCW polariton amplitudes determine both directional accessibility and the coherent distribution of a single photon between the two propagation paths. In the driven-dissipative steady state, this common microscopic structure allows strong directional photon blockade to coexist with substantial single-photon path entanglement. More strikingly, switching between the two one-photon resonance branches reverses both the photon-statistical and intensity asymmetries while approximately preserving the corresponding conditional path concurrence under exchange of the input direction. The persistence of this correspondence over broad ranges of Zeeman splitting and backscattering shows that directional photon statistics and single-photon path entanglement, although rooted in the same polariton structure, remain distinct aspects of the quantum response.

The underlying mechanism should also extend beyond the minimal ring-cavity setting considered here. Combining path complementarity with chiral light-matter interfaces could enable directional quantum nodes that route nonclassical photons while retaining coherent path encoding~\cite{lodahl2017chiral,mahmoodian2016quantum, ramos2016non}. Combining this mechanism with nonreciprocal or chiral photon blockade could further couple control of propagation, photon statistics, and path coherence within a common quantum-optical architecture~\cite{Huang18,PhysRevLett.134.013602,PhysRevX.5.021025,zheng2025nonreciprocal}. In coupled resonators or emitter arrays, path complementarity may acquire a collective character through polaritonic bands and many-body correlations~\cite{hartmann2008quantum,tang2022nonreciprocal}. More broadly, these directions point toward quantum-light interfaces that combine nonclassical photon statistics, directional transport, and path-encoded quantum information~\cite{luxmoore2013interfacing,caspar2020heralded,Xiao2021chiral}. Such hybrid quantum interfaces may offer new opportunities for
direction-selective quantum communication, photonic information
processing, and programmable quantum optical devices~\cite{wang2020integrated,elshaari2020hybrid,madsen2022quantum}.

{\em Acknowledgments}.---This work was supported by the National Natural Science Foundation of China (Grant No.12374365 and Grant No. 12674338), Quantum Science and Technology-National Science and Technology Major Project (Grant No.2025ZD0300400), Guangdong Provincial Quantum Science Strategic Initiative (Grants No. GDZX2505001 and No. GDZX2604004), and Guangdong University of Technology SPOE Seed Foundation (SF2024111504).

%\bibliography{ref_noncip_PB}


\end{document}